\documentclass[sn-mathphys,Numbered]{sn-jnl}
\usepackage{graphicx}%
\usepackage{multirow}%
\usepackage{amsmath,amssymb,amsfonts}%
\usepackage{amsthm}%
\usepackage{mathrsfs}%
\usepackage[title]{appendix}%
\usepackage{xcolor}%
\usepackage{textcomp}%
\usepackage{manyfoot}%
\usepackage{booktabs}%
\usepackage{algorithm}%
\usepackage{algorithmicx}%
\usepackage{algpseudocode}%
\usepackage{listings}%
\usepackage{float}
\usepackage{subcaption}
\usepackage[T1]{fontenc}
\usepackage{lmodern} 
\usepackage{soul} 
\usepackage{xcolor} 
\definecolor{lightyellow}{rgb}{1, 1, 0.88} 
\sethlcolor{lightyellow} 
\begin{document}
\title[Article Title]{NOMA-CSK Integrated VLC System with Reinforcement Learning-Based Multi-Objective Power Allocation}
\author*[1]{\fnm{Serkan} \sur{Vela}}\email{serkanvela@ktu.edu.tr}
\author[2]{\fnm{Gokce} \sur{Hacioglu}}\email{gokcehacioglu@ktu.edu.tr}
\affil*[1]{\orgdiv{Electronics and Communication Engineering}, \orgname{Karadeniz Technical University}, \orgaddress{\street{Irfanli}, \city{Trabzon}, \postcode{61830}, \country{Turkey}}}
\affil[2]{\orgdiv{Electrical and Electronics Engineering}, \orgname{Karadeniz Technical University}, \orgaddress{\street{Universite}, \city{Trabzon}, \postcode{61080}, \country{Turkey}}}

\abstract{This paper introduces a novel framework that synergistically combines Non-Orthogonal Multiple Access (NOMA) with Color Shift Keying (CSK) modulation to substantially boost spectral efficiency in Visible Light Communication (VLC) systems. A key challenge in the proposed NOMA-CSK architecture is managing the complex power allocation process, especially under cross-color interference caused by spectral overlap among LEDs and the limitations of optical filters. To overcome this, we develop an intelligent power allocation strategy powered by a Soft Actor-Critic (SAC) reinforcement learning agent. Trained in a simulated indoor environment, the SAC agent dynamically distributes power among users with diverse channel conditions while balancing multiple performance objectives. Simulation results show that our SAC-based method significantly outperforms traditional approaches such as Gain Ratio Power Allocation (GRPA) and Normalized Gain Difference Power Allocation (NGDPA), achieving superior fairness, higher overall throughput, and reduced bit error rates—even under a challenging 10 dB SNR. Notably, the trained agent demonstrates strong generalization capabilities, maintaining optimal performance in unseen environments without requiring retraining. Overall, this work makes two major contributions: it presents a pioneering NOMA-CSK VLC system design and delivers a robust, adaptive power allocation solution critical for real-world applications.}

\keywords{NOMA, CSK, power allocation, RL, SAC agent}
\maketitle
\section{Introduction}\label{sec1}
With the development and widespread use of wireless communication systems, the number of wireless devices and data transfer requirements are increasing. However, the current frequency band of 6 GHz and below has become too crowded to provide efficient communication \cite{cisco2020}. Therefore, it is suggested to use the wide unlicensed spectrum of visible light in wireless communication systems \cite{rajagopal2012ieee}. Visible Light Communication (VLC) systems offer a number of benefits, including energy savings due to the ability to carry out lighting and communication simultaneously, security, privacy, and resistance to electromagnetic radiation \cite{pathak2015visible,armstrong2013visible}.

VLC systems usually utilize light-emitting diodes (LEDs) as transmitters, while photodetectors (PDs) are favored as receivers due to their quick response times. The modulation speeds of LEDs are a crucial element in communication performance. While phosphor-coated white LEDs have modulation speeds of approximately 10 MHz, multicolored LEDs, such as Red, Green, Blue (RGB) and Quadrature LED (QLED), can be modulated at speeds up to 100 MHz. Therefore, compared to white LEDs with phosphor coating, communication via multicolored LEDs may theoretically be possible at higher speeds.

In wireless communication systems, the modulation type, which is one of the factors affecting communication performance, is as important as data rates. Various modulation methods such as On-Off Keying (OOK) \cite{tao2018power}, Pulse Width Modulation (PWM) \cite{basha2019design}, and Color Shift Keying (CSK) \cite{rajagopal2012ieee} can be used as modulation types in VLC systems. The newest of these, CSK, is a modulation method published in the IEEE 802.15.7 standard and used in VLC systems with multi-colored LED transmitters \cite{rajagopal2012ieee}. The CSK modulation method is a modulation method in which the color of the light emitted from the transmitter is changed according to the information signal in multicolor LED VLC systems. On the receiver side, there are color filters. Signals obtained from color filters are used in demodulation. For RGB LED emitter systems, the IEEE 802.15.7 standard defines 4-CSK, 8-CSK, and 16-CSK.

CSK has several advantages in VLC systems. Firstly, it maintains a constant power envelope of the transmitted signal, which reduces the possibility of health issues caused by fluctuations in light intensity. Secondly, it provides color lights with a stable temperature. Thirdly, with rapid switching, the flicker effect is minimized. Fourthly, even at slow switching speeds, high-level modulation is possible, which increases the capacity \cite{rajagopal2012ieee}. Additionally, CSK does not require additional DC power for data transmission, giving it an edge over other modulation schemes utilized in VLC. This allows for an increase in the signal-to-noise ratio, and efficient use of the entire transmission spectrum is possible. CSK modulation involves instantaneously changing the color of light to transmit data. As a result, illuminance \cite{dowhuszko2021effect, gancarz2013impact} can change during communication.

Besides CSK modulation, multiple access methods in VLC have also been extensively studied in the literature. Both orthogonal and non-orthogonal multiple access methods (OMA and NOMA) are available in VLC. Power domain NOMA (PD-NOMA), which is a non-orthogonal multiple access method, has been proposed as a promising and bandwidth-efficient solution for multiple access in indoor VLC networks \cite{marshoud2015non,marshoud2017performance}. NOMA works by superimposing user signals in the power domain, where a separate power level is assigned to each signal based on the user's channel conditions. In NOMA, all users have continuous access to the full bandwidth. The fundamental tenet of NOMA is to give users with weak channels more power than those with strong channels. To achieve this, the user with the highest power can detect the information signal despite the interference caused by other users, while other users perform successive interference cancellation (SIC) to decode their own signals \cite{manglayev2017noma}. It has been demonstrated in \cite{kizilirmak2015non} that NOMA can achieve higher data rates than Orthogonal Frequency Division Multiple Access (OFDMA) in VLC systems. Similarly, in \cite{yin2015performance}, researchers examined the performance of indoor VLC systems using NOMA and discovered a substantial enhancement in system capacity due to the implementation of NOMA. In \cite{tsiropoulou2016resource}, NOMA techniques were compared with classical OFDMA techniques, and it was found that some of the advantages of NOMA include better quality of service, less interference, and the ability to provide multiple access to a larger number of users.

Power allocation is a critical challenge in NOMA-VLC systems as it directly determines the achievable capacity and user fairness \cite{vela2019tr,vela2024novel}. The gain ratio power allocation (GRPA) method was proposed in \cite{marshoud2015non} for a scenario with two transmitters and three receivers, and it showed better sum-rate and Bit Error Rate (BER) than static power allocation in the best-case scenario. In \cite{yin2016performance}, the normalized gain difference power allocation (NGDPA) method was introduced and compared to GRPA, resulting in a 29.1\% higher total capacity in a 3-user 2x2 MIMO VLC system. An optimization algorithm was proposed in \cite{yang2021power} to maximize the overall data rate and compared to GRPA. However, fairness and Quality of Service (QoS) were not considered. In \cite{tahira2019optimization}, a convex decoder was used to optimize the total data rate and BER under a certain light intensity constraint, but users' data rate and fairness were not considered. The VOOK method with GRPA was used in \cite{tao2018power} for power allocation in NOMA-VLC systems with variable illumination levels. However, VOOK is inefficient as the total capacity decreases in proportion to brightness. In \cite{eltokhey2021power}, a particle swarm optimization (PSO)-based power allocation optimization algorithm for NOMA VLC multicellular networks was proposed, but the level of fairness achieved was less than 0.6, which is suboptimal for QoS. When distributing LED transmitter power to users in VLC systems, factors such as fairness, illumination level, and user data rate should be taken into account to provide eye comfort and better QoS.

Recent research has explored the use of reinforcement learning (RL) methods to optimize communication systems. Using deep reinforcement learning (DRL) as the method, Xiao et al. \cite{xiao2017reinforcement} investigated power allocation for non-orthogonal multiple access (NOMA) against smart jammers in radio frequency (RF) systems, resulting in an increased sum-rate. He et al. \cite{he2019joint}, on the other hand, studied power distribution in multi-carrier RF systems with NOMA and observed an overall system performance improvement. Zhang et al. \cite{zhang2019energy} employed DRL to investigate energy-efficient resource allocation, while Zhang et al. \cite{zhang2020deep} demonstrated a 32.9\% higher efficiency compared to the ALOHA method. Wang et al. \cite{wang2021joint} proposed the DRL-JRM (Deep Reinforcement Learning-Joint Resource Management) scheme for RF multi-carrier systems in NOMA, which showed superior system efficiency and resistance to interference, especially in the presence of large numbers of users and strong intercellular interference. Bariah et al. \cite{bariah2022deep} investigated Deep Q-Learning-Based resource allocation to increase the energy efficiency and sum-rate for NOMA VLC systems, showing its superiority over the results obtained by genetic algorithm and differential evolution. Al Hammadi et al. \cite{al2022joint} proposed a Deep Q-Learning-based approach for resource allocation in NOMA Visible Light Communications (VLC), aiming to enhance the system's efficiency and sum-rate. Finally, Al et al. \cite{al2022joint} investigated Q-Learning-Based resource allocation and LED emitter angle optimization to increase sum-rate for NOMA VLC systems.

While NOMA and CSK have been investigated independently for VLC, their integration to leverage the benefits of both power-domain multiplexing and color-based modulation remains an unexplored area. This combination presents a unique opportunity to maximize user capacity and throughput, but it also introduces a more complex resource allocation problem. Existing power allocation schemes like GRPA and NGDPA were not designed for the specific interference characteristics of a NOMA-CSK system. It is important to note that these conventional methods were primarily designed for NOMA systems where interference could be modeled more simply, and not for the complex cross-color interference inherent in our proposed NOMA-CSK architecture. Particularly, the complex interference patterns arising from the spectral overlap of LEDs and non-ideal optical filters lead to significant cross-color interference. This complex environment makes it challenging for traditional analytical models to find an optimal and fair power allocation, necessitating an adaptive, data-driven approach. While traditional optimization methods can find solutions for well-defined, static channel models, they struggle to adapt to the dynamic and stochastic nature of real-world VLC environments with complex, non-linear cross-color interference. Reinforcement learning, particularly the SAC agent, offers a powerful alternative by learning a robust control policy directly from environmental interactions, enabling it to generalize to unseen channel conditions without requiring an explicit analytical model of the complex interference.

To address this research gap, this paper makes a twofold contribution. First, we propose and model a novel PD-NOMA VLC system that utilizes CSK modulation (NOMA-CSK). Second, we solve the inherent power allocation challenge within this new system by developing a fairness-driven SAC reinforcement learning agent. Our key contributions are:
\begin{itemize}
\item \textbf{A Novel NOMA-CSK System Architecture:} We design and model a VLC system that combines power-domain NOMA with CSK, establishing a framework for efficient multi-user communication.
\item \textbf{A Fairness-Centric Multi Ojective RL Power Allocation Solution:} Within this new system, we design and train a SAC agent with a reward mechanism that focuses on Jain's fairness index,ensuring equitable resource distribution, along with the user rates, bit error rates, power efficiency and stability.
\item \textbf{Superior Performance and Robustness:} We demonstrate through extensive simulations that our RL-based approach significantly outperforms traditional benchmarks in sum-rate, BER, and fairness. Crucially, we prove the agent's ability to generalize, as it maintains high performance in unseen environments without retraining, underscoring its practical value.
\end{itemize}

This work lays the foundation for NOMA-CSK as a viable technology and provides the critical intelligent mechanism needed for its efficient operation.

\section{VLC Channel}\label{sec2}

The VLC channel can be modeled in two ways: line of sight (LoS) and non-line of sight (NLoS). According to \cite{zeng2009high}, the strongest NLoS channel of the VLC channel is 7dB weaker than the weakest LoS channel. Therefore, this study is based on the LoS model. According to the model in \cite{lee2011indoor}, the attenuation of light is established for the case of direct vision between the transmitter and receiver and its mathematical expression is given in Eq.\ref{eqn:vlc}.

\begin{equation}
H_{LoS}=\frac{A_r R(\phi_{T_{x}})}{d^2} T_S(\psi) g(\psi) \cos(\psi) \label{eqn:vlc}
\end{equation}

where $R(\phi_{Tx})=\frac{(k_l+1)}{2\pi}\cos^{k_l}(\phi_{Tx})$ is the LED's radiant intensity, $k_l=(-\ln2)/\ln(\cos(\Omega_{1/2}))$ is the Lambertian emission mode number, $d$ is the distance between transmitter and receiver, $\phi_{Tx}$ is the radiation angle of the LED, $\Omega_{1/2}$ is the LED half angle, $A_r$ is the surface area of the photodiode (PD), $\psi$ is the incidence angle of the PD, $T_s(\psi)$ is the optical filter gain, $g(\psi)=n^2/\sin^2\phi$ is the optical lens gain, $n$ is the refractive index of the PD, and $\phi$ is the view angle of the PD.

It is important to note that this study primarily considers the Line-of-Sight (LoS) transmission model, which constitutes the dominant signal component in most indoor VLC scenarios. While this is a common and valid simplification for tractable analysis, real-world environments also include Non-Line-of-Sight (NLoS) components arising from reflections. These reflections could alter the channel characteristics and interference profile. The investigation of NLoS effects on the performance of the NOMA-CSK system and the adaptability of the SAC agent to such conditions remains a valuable direction for future work.

\section{System Model}
The simulations are conducted in a virtual office space measuring 5 meters in length, 5 meters in width, and 3 meters in height. The OSRAM LZ4-00MA00 multicolor LED, which is commercially available, has been selected for this particular application. We assumed that the emitted lighting powers for the red (R), green (G), and blue (B) colors are equivalent.

LEDs can be represented in various ways for simulation purposes, and the H model proposed in \cite{he2010model} closely approximates the results obtained from experimental studies. The H model parameters for LZ4-00MA00 LED can be found in Table \ref{tab:1} \cite{ge2018optical}, and these parameters are utilized to compute the Power Spectral Density (PSD) (Eq.\ref{eqn:6}) of the illuminating LED by using,Eq.\ref{eqn:7} and Eq.\ref{eqn:8}.

\begin{equation}
g(\lambda,\lambda_p,\Delta\lambda)=\exp{\left[-\frac{(\lambda-\lambda_p)^2}{\Delta\lambda^2}\right]} \label{eqn:6}
\end{equation}
\begin{equation}
\Delta\lambda=
\begin{cases}
    \Delta\lambda_1, & \lambda < \lambda_p \\
    \Delta\lambda_2, & \lambda \geq \lambda_p
\end{cases}
\label{eqn:7}
\end{equation}
\begin{equation}
PSD_i(\lambda)=\frac{g(\lambda,\lambda_p,\Delta\lambda)+k_1g^{k_2}(\lambda,\lambda_p,\Delta\lambda)}{1+k_1}
\label{eqn:8}
\end{equation}

Where $\lambda_p$ is the peak wavelength, $\Delta\lambda_1$ and $\Delta\lambda_2$ are the left/right half spectral widths, and $k_1$, $k_2$ are characteristic shape parameters for color channel $i$. 

\begin{table}[h!]
\centering
\caption{Parameter values for different color components \cite{ge2018optical}}\label{tab:1}
\begin{tabular}{@{}cccc@{}}
\toprule
Parameter & Red & Green & Blue \\
\midrule
$\lambda_p$ & 632.5 & 517.7 & 453 \\
$\Delta\lambda_1$ & 23.84 & 29.38 & 18.99 \\
$\Delta\lambda_2$ & 14.74 & 45.21 & 25.5 \\
$k_1$ & 2 & 2 & 2 \\
$k_2$ & 6 & 3 & 5 \\
\bottomrule
\end{tabular}
\end{table}
\begin{figure}[h!]
\centering
\begin{subfigure}[b]{0.56\textwidth}
\includegraphics[width=\textwidth]{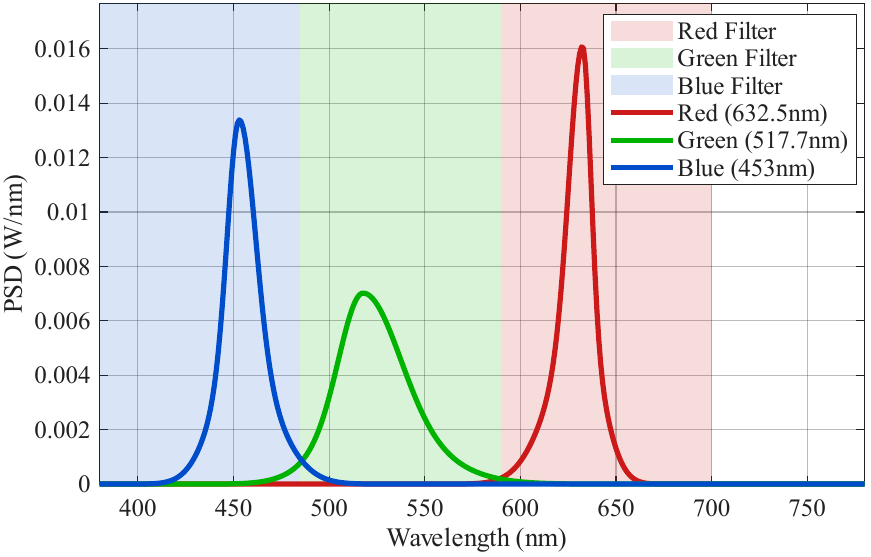}
\caption{Power spectra of the LZ4-00MA00 without amber color}
\label{fig:1a}
\end{subfigure}
\hfill
\begin{subfigure}[b]{0.43\textwidth}
\includegraphics[width=\textwidth]{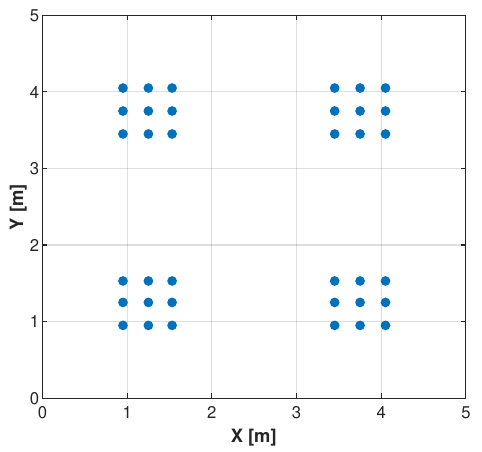}
\caption{Locations of LED transmitters}
\label{fig:2a}
\end{subfigure}
\caption{Power spectra and LED transmitter locations}
\label{fig:1-2}
\end{figure}

The R, G, and B spectra's intensity changes when the data to be transmitted is modulated using CSK modulation. Color filters can be used at the receiver to identify changes in these intensities. In this instance, Eq.\ref{eqn:9} describes the power of the signals that were received from the $j$-th color filter.

\begin{equation}
p_{i,j}^{(k)}=\frac{1}{3} \int_{\lambda_l^i}^{\lambda_h^i} P S D(\lambda) F_j(\lambda) S_i h^{(k)} d \lambda ; i, j \in \{1,2,3\}
\label{eqn:9}
\end{equation}

In the context provided, $i$ and $j$ correspond to the colors of the LED at the transmitter and the photodetector at the receiver, respectively. The values of $i$ and $j$ range from $1$ to $3$, representing the colors red, green, and blue. The variables $\lambda_l^{i}$ and $\lambda_h^{i}$ denote the lower and upper wavelengths of the spectrum for the LED color $i$, respectively. The variable $F_{j}$ represents the gain associated with the $j$-colored optical filter. The variable $S_{i}$ denotes the CSK symbol value for color $i$, and finnaly $h$ is the VLC channel attenuation. The matrix $\boldsymbol{p}$ is a $3\times3$ channel matrix, where the $(i,j)$-th element is determined by Eq. \ref{eqn:9}.
\begin{equation}
\boldsymbol{p}=\begin{bmatrix}p_{1,1}&p_{2,1}&p_{3,1}\\p_{1,2}&p_{2,2}&p_{3,2}\\p_{1,3}&p_{2,3}&p_{3,3}\end{bmatrix}
\label{eqn:10}
\end{equation}
The non-ideal nature of both the optical filters and the LEDs results in the matrix $\boldsymbol{p}$ not being a diagonal matrix. The elements outside the diagonal reflect the power of interference. In this case, the SINR at the receiver is:
\begin{equation}
\operatorname{SINR}=\frac{\sum_{i,j} p_{i, j} \delta_{i j}}{\sum_{i,j} p_{i,j} - \sum_{i,j} p_{i, j} \delta_{i j}+N_0}
\label{eqn:11}
\end{equation}
Here, $\delta_{i j}$ is the Kronecker delta function and $N_0$ is the power of AWGN noise respectively. Capacity calculation over SINR can be done according to Eq.\ref{eqn:12}. After the CSK demodulation is performed on the received signal, the throughput can be calculated using Eq.\ref{eqn:13}.

\begin{equation}
C=B\log_2(1+\text{SINR}) \label{eqn:12}
\end{equation}
\begin{equation}
T= C(1-\text{BER}) \label{eqn:13}
\end{equation}

The open-source program shared in \cite{matlabexchange_pspectro} can be used to calculate the illumination metrics such as CRI \cite{gancarz2013impact}, CCT \cite{dowhuszko2021effect}, and luminous flux. The irradiance data corresponding to the LED's wavelength is input into the program, which then uses it to calculate the illumination parameters. The CRI, CCT, and illumination values should be calculated over the long-term average of the irradiance value because the brightness of the light emitted from the R, G, and B colors varies with communication.

\section{Numerical Results}
In this section, we simulated the environment to evaluate the performance of the proposed NOMA-CSK system for VLC networks. The variables associated with the VLC channel, along with simulation parameters, are detailed in Table \ref{tab:2}. Our simulation setup includes a NOMA-CSK communication system with two receivers ($U_1$ and $U_2$), each equipped with three photodetectors (PDs) positioned perpendicular to the ceiling.

\begin{table}[h!]
\centering
\caption{VLC channel and simulations parameters}\label{tab:2}
\begin{tabular}{@{}ll@{}}
\toprule
\textbf{Parameter} & \textbf{Value} \\
\midrule
RGB LED viewing angle & $60^\circ$ \\
Power of an RGB LED & 1 W \\
PD responsivity & 0.54 A/W \\
PD surface area & $1~\mathrm{cm}^2$ \\
Optical ideal filter lower cutoff wavelengths & [590, 485, 380] [R,G,B] nm \\
Optical ideal filter upper cutoff wavelengths & [700, 590, 485] [R,G,B] nm \\
Optical filter gain & 1 \\
Optical lens gain & 1 \\
Transmitter-receiver bandwidth & 30 MHz \\
SNR & 10 dB \\
Modulation Level & 4 \\
\bottomrule
\end{tabular}
\end{table}

Both users have three PDs with suitable color filters in front of each PD for the VLC-NOMA-CSK simulation. The following Eq.\ref{eqn:14} describes how NOMA is applied to the CSK VLC system:
\begin{equation}
p_{i, j}^{(k)}=\frac{1}{3} \int_{\lambda_l^i}^{\lambda_h^i} P S D(\lambda) F_j(\lambda)\left[S_i^{(1)} {\rho^{(1)}}+S_i^{(2)} {(1-\rho^{(1)})}\right] h^{(k)} d \lambda
\label{eqn:14}
\end{equation}


Here, $S_i^{(1)}$ and $S_i^{(2)}$ are the value of the CSK symbol in color $i$ for $U_1$ and $U_2$, respectively. $\rho^{(1)}$ is the NOMA power allocation coefficient for $U_1$, who has a better channel gain ($h^{(1)} > h^{(2)}$). The total channel gain between the transmitter and the $k$-th user is $h^{(k)}$.

The received power component at user $k$ originating from the signal intended for user $u \in \{1, 2\}$ is given by:
\begin{equation}
P_{i,j}^{(k, u)} = \frac{1}{3} \int_{\lambda_l^i}^{\lambda_h^i} PSD(\lambda) F_j(\lambda) S_i^{(u)} \rho^{(u)} h^{(k)} d\lambda
\label{eq:power_component}
\end{equation}
where $\rho^{(2)} = 1 - \rho^{(1)}$.

The total desired signal power for user 1 ($U_1$) and user 2 ($U_2$) are:
\begin{equation}
S^{(1)} = \sum_{i,j} P_{i,j}^{(1, 1)} \delta_{ij} \quad \text{and} \quad S^{(2)} = \sum_{i,j} P_{i,j}^{(2, 2)} \delta_{ij}
\end{equation}

The cross-color interference for user $k$ from its own signal is $I_{CCI}^{(k)} = \sum_{i,j} P_{i,j}^{(k, k)} (1-\delta_{ij})$.

With Successive Interference Cancellation (SIC), user 1 first decodes and removes user 2's signal. However, since user 2 has higher power, its signal acts as interference for user 1. Conversely, user 2 sees user 1's signal as interference. The inter-user interference experienced by user 2 is $I^{(2 \leftarrow 1)} = \sum_{i,j} P_{i,j}^{(2, 1)}$.

The SINR at the receivers can now be expressed unambiguously:
\begin{equation}
\operatorname{SINR}^{(1)}=\frac{S^{(1)}}{I_{CCI}^{(1)} + \sum_{i,j} P_{i,j}^{(1, 2)} + N_0^{(1)}}
\label{eqn:15}
\end{equation}
\begin{equation}
\operatorname{SINR}^{(2)}=\frac{S^{(2)}}{I_{CCI}^{(2)} + I^{(2 \leftarrow 1)} + N_0^{(2)}}
\label{eqn:16}
\end{equation}

$N_0^{(1)}$ and $N_0^{(2)}$ represent the AWGN noise powers for $U_1$ and $U_2$ respectively. $I^{(1)}$ denotes the interference caused by the message signals of $U_1$ on $U_2$, and it can be expressed as follows:

\begin{equation}
I^{(1)}=\sum_i \sum_j\left(\int_{\lambda_l^i}^{\lambda_h^i} P S D(\lambda) F_j S_i^{(1)} \rho^{(1)} h^{(1)} \mathrm{~d} \lambda\right)
\label{eqn:18}
\end{equation}

In a practical communication system, the positions of users are not fixed, and the channel conditions change as they move. These factors have been considered in the proposed reinforcement learning-based NOMA-CSK power allocation method. Our objective was to train the agent to efficiently allocate power even when confronted with channel conditions that differ from those encountered during training.

To provide a clear baseline for comparison, we detail the established power allocation schemes used as benchmarks. It is crucial to reiterate that these methods were developed for scenarios without the complex cross-color interference present in our NOMA-CSK model, but we implement them here to establish a performance benchmark.

\textbf{GRPA-NOMA}: This method allocates power proportionally to the square of the channel gains between users. The powers allocated to users 1 and 2 ($P_1, P_2$) are determined as follows, assuming user 1 has the better channel condition ($h^{(1)} > h^{(2)}$):
\begin{gather}
{P_1 = \frac{P_{total}}{1+\frac{(h^{(1)})^2}{(h^{(2)})^2}}} \label{eq:grpa_p1} \\
P_2 = P_{total} - P_1 \label{eq:grpa_p2}
\end{gather}

Here, $P_{total}$ is the total transmission power. The power allocation coefficient for user 1 is $\rho^{(1)} = P_1 / P_{total}$.

\textbf{NGDPA-NOMA}: This method considers the normalized difference between channel gains to determine power allocation. The power allocation is given by:
\begin{gather}
P_1 = \frac{P_{total}}{1+\frac{h^{(1)}}{|h^{(1)}-h^{(2)}|}} \label{eq:ngdpa_p1} \\
P_2 = P_{total} - P_1 \label{eq:ngdpa_p2}
\end{gather}
For both methods, the resulting throughput for each user is calculated based on their respective SINR values as defined in Eq. \ref{eqn:15} and \ref{eqn:16}.

\textbf{Reinforcement Learning}: Reinforcement learning is a method for training an agent in a simulation environment through trial and error, without relying on a pre-existing dataset. During these trials, the agent learns a policy that maximizes rewards by receiving a positive reward when it achieves a desired outcome, and a negative reward (penalty) otherwise. In this study, a model-free, online, off-policy, actor-critic reinforcement learning agent SAC is utilized. Since both the action and observation values are continuous, an agent capable of working in this environment is selected as the SAC agent \cite{haarnoja2018soft}.

The environment should be designed before the learning process can be implemented. The actions that the agent can take and the quantities it can observe should be determined to design the environment. 

To formulate the problem within the RL framework, we define the state, action, and reward function.

\begin{itemize}
    \item \textbf{State (Observation) Space ($\mathcal{S}$):} The agent observes the state of the channel, which is captured by the ratio of the users' channel gains. The state $s_t$ at time step $t$ is:
    \begin{equation}
    s_t = \frac{h^{(2)}}{h^{(1)}} \in [0, 1]
    \label{eq:observation_space}
    \end{equation}
    where $h^{(1)}$ and $h^{(2)}$ are the channel gains of the stronger and weaker users, respectively.

    \item \textbf{Action Space ($\mathcal{A}$):} The agent's action, $a_t$, is to select the NOMA power allocation coefficient $\rho^{(1)}$ for the user with the better channel gain. To ensure the weaker user receives more power, this coefficient is bounded.
    \begin{equation}
    a_t = \rho^{(1)} \in [0, 0.5]
    \label{eq:action_space}
    \end{equation}
    The power for the second user is then $1 - \rho^{(1)}$.

    \item \textbf{Reward Function ($R_t$):} The reward function guides the agent's learning process. It is a multi-objective function designed to maximize fairness and overall performance, as detailed in Eq. \ref{eq:total_reward}.
\end{itemize}


The users' channel gains are determined to fall within the range of $2.84 \times 10^{-5}$ and $5.98 \times 10^{-4}$. This range encompasses the maximum and minimum values among 2601 channels, obtained by varying the users' coordinates as $0.5 < X < 4.5$, $0.5 < Y < 4.5$, and $0.5 < Z < 2.5$, with an increment of 0.25 for each step. The values of $h^{(1)}$ and $h^{(2)}$ are generated randomly according to Eq.\ref{eqn:19} and Eq.\ref{eqn:20}, respectively.

\begin{equation}
2.84\times{10}^{-5} < h^{(1)} < \frac{5.98\times{10}^{-4}}{r}
\label{eqn:19}
\end{equation}
\begin{equation}
h^{(2)} = h^{(1)} \times r
\label{eqn:20}
\end{equation}

After exploring the combination of NOMA and CSK for VLC systems, the subsequent critical step involves delineating the agent's performance criteria and configuring its reward mechanism. This pivotal phase is essential for guiding the training regimen, which is facilitated by the SAC algorithm, as delineated in Algorithm \ref{algo}. SAC is renowned for its effectiveness in addressing complex decision-making tasks.

Central to the SAC algorithm's efficacy is the precise calibration of various parameters, detailed in Table \ref{tab:rl_parameters}. The hyperparameters in Table \ref{tab:rl_parameters} were selected based on common practices in DRL literature and refined through preliminary experiments. 

\begin{table}[ht!]
\centering
\caption{RL Parameters}\label{tab:rl_parameters}
\begin{tabular}{@{}lc@{}}
\toprule
\textbf{Parameter} & \textbf{Value} \\
\midrule
Fairness observation space $O$ & $[0,1]$ \\
Power allocation action space $A$ & $[0,0.5]$ \\
Discount factor $\gamma$ & 0.99 \\
Entropy coefficient $\alpha$ & 1 \\
Target smoothing coefficient $\tau$ & $10^{-3}$ \\
Learning rate $\eta$ & $3\times10^{-4}$ \\
Replay buffer size $B$ & $10^4$ \\
Mini-batch size $M$ & 64 \\
MaxEpisodes & 1000 \\
MaxStepsPerEpisode & 500 \\
StopTrainingCriteria & AverageReward \\
ScoreAveragingWindowLength & 100 \\
StopTrainingValue & 163 \\
\bottomrule
\end{tabular}
\end{table}




The agent's training is guided by a multi-objective reward function, $R_t$, designed to balance fairness, throughput, and system stability at each time step $t$. The total reward is a summation of several components:

\begin{equation}
R_t = R_{\text{fair}} + R_{\text{sum-rate}} + R_{\text{BER}} + R_{\text{power}} + R_{\text{stability}}
\label{eq:total_reward}
\end{equation}

Each component is defined as follows:

\begin{itemize}
    \item \textbf{Fairness Reward ($R_{\text{fair}}$):} To enforce equitable resource distribution, a significant reward is given for achieving near-perfect fairness, with a quadratic penalty for deviations. Jain's fairness index, $J$, is calculated from the throughputs $T_k$ of $N$ users as defined in Eq. \ref{eq:jain_index}.
    \begin{equation}
    J = \frac{\left(\sum_{k=1}^N T_k\right)^2}{N \sum_{k=1}^N T_k^2}
        \label{eq:jain_index}
    \end{equation}
    The fairness reward is then calculated as:
    \begin{equation}
    R_{\text{fair}} = 
    \begin{cases} 
    +100, & \text{if } J \geq 0.99 \\
    -50 \times (0.99 - J)^2, & \text{otherwise}
    \end{cases}
    \label{eq:reward_fairness}
    \end{equation}

    \item \textbf{Sum-Rate Bonus ($R_{\text{sum-rate}}$):} To encourage overall performance, a bonus is given that is proportional to the throughput value in Mbps, which is obtained by dividing the total system throughput in bps (bits per second) by $10^8$.
    \begin{equation}
    R_{\text{sum-rate}} = w_{sr} \left( \frac{\sum_{k=1}^{N} T_k}{10^8} \right)
    \label{eq:reward_sumrate}
    \end{equation}
    Here, $T_k$ represents the throughput of the $k$-th user in bps, and the weight coefficient $w_{sr}$ is set to 0.3.

    \item \textbf{BER Penalty ($R_{\text{BER}}$):} A penalty is applied if the BER for any user exceeds a predefined threshold, ensuring communication reliability.
    \begin{equation}
    R_{\text{BER}} = 
    \begin{cases} 
    -20, & \text{if } \text{BER}_k > 10^{-3} \text{ for any user } k \\
    0, & \text{otherwise}
    \end{cases}
    \label{eq:reward_ber}
    \end{equation}

    \item \textbf{Power Efficiency Bonus ($R_{\text{power}}$):} This component encourages the agent to select power allocation coefficients ($\rho$) that are not at the extremes, promoting a balanced allocation.
    \begin{equation}
    R_{\text{power}} = 10 \times (1 - |\rho^{(1)} - 0.25|)
    \label{eq:reward_power}
    \end{equation}

    \item \textbf{Stability Bonus ($R_{\text{stability}}$):} A small reward is given for operating within a stable power allocation range, penalizing selections at the absolute boundaries.
    \begin{equation}
    R_{\text{stability}} = 
    \begin{cases} 
    +5, & \text{if } 0.01 \leq \rho^{(1)} \leq 0.49 \\
    -10, & \text{otherwise}
    \end{cases}
    \label{eq:reward_stability}
    \end{equation}
\end{itemize}

An episode terminates successfully if the fairness target is met, yielding an additional bonus of +50. Conversely, failure conditions result in a penalty of -100.


The SAC agent is trained for 1000 episodes by randomly generating user channel ratios and channels. Training is completed when the average reward over the last 100 episodes is 163. The agent is then tested with 1000 randomly generated pairs of $(h^{(1)}, h^{(2)})$ where the value of $r$ is varied in increments of 0.001 within the range of 0 to 1.

\begin{figure}[h!]
\centering
\begin{subfigure}[b]{0.49\textwidth}
  \includegraphics[width=\textwidth]{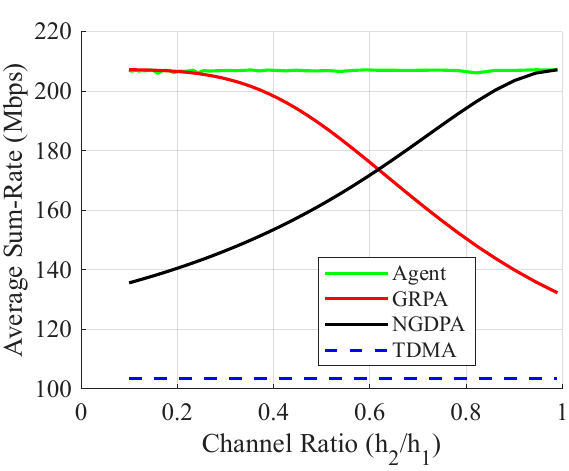}
\caption{Sum-rate performance}
\label{fig:3a}
\end{subfigure}
\hfill
\begin{subfigure}[b]{0.49\textwidth}
\includegraphics[width=\textwidth]{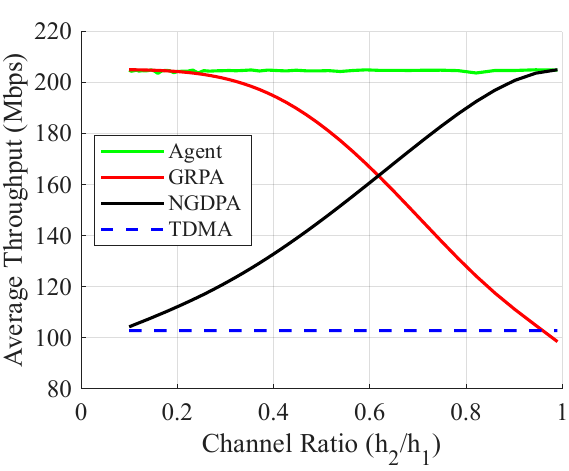}
\caption{Throughput performance}
\label{fig:4a}
\end{subfigure}
\vskip\baselineskip
\begin{subfigure}[b]{0.49\textwidth}
\includegraphics[width=\textwidth]{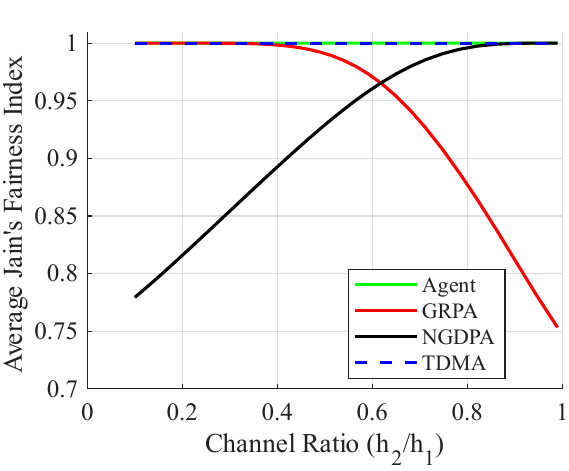}
\caption{Fairness performance}
\label{fig:5a}
\end{subfigure}
\hfill
\begin{subfigure}[b]{0.49\textwidth}
\includegraphics[width=\textwidth]{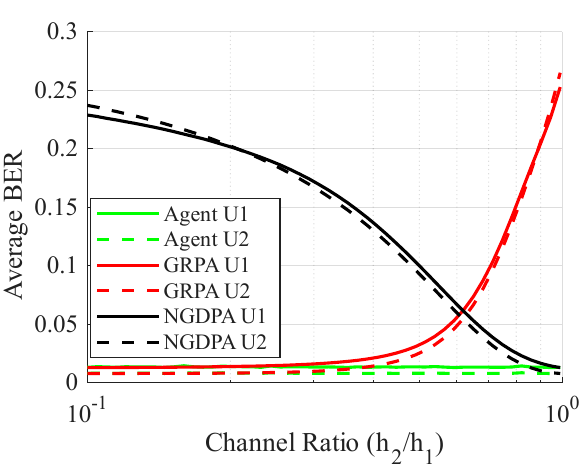}
\caption{BER performance}
\label{fig:6a}
\end{subfigure}
\caption{Performance comparison of the proposed SAC agent against GRPA, NGDPA, and TDMA in the trained environment.}
\label{fig:3-6}
\end{figure}

The changes in sum-rate, throughput, fairness, and BER values are depicted in Fig.~\ref{fig:3a}, Fig.~\ref{fig:4a}, Fig.~\ref{fig:5a} and Fig.~\ref{fig:6a} respectively, as the channel ratio $r$ varies. These values were obtained by applying proposed power allocation techniques to users in randomly generated $(h^{(1)}, h^{(2)})$ channels. The performance curves for the SAC agent in the figures represent the average results obtained from 500 evaluation runs with different random seeds, confirming the stability and reliability of the learned policy. The figures also include the performance results of TDMA-based multiple access, GRPA, and NGDPA. Upon examination, it becomes clear that the trained agent demonstrates significantly superior performance in sum-rate, throughput, and fairness compared to other methods.

\begin{algorithm}[h!]
\caption{SAC Agent Training for NOMA-CSK Power Allocation}\label{algo}
\begin{algorithmic}[1]
\State \textbf{Input:} State space $\mathcal{S}$, Action space $\mathcal{A}$; parameters $\gamma, \alpha, \tau, \eta$; buffer size $B$; batch size $M$.
\State \textbf{Initialize:} Actor network $\pi_\phi$, two critic networks $Q_{\theta_1}, Q_{\theta_2}$, and target networks $Q'_{\theta'_1}, Q'_{\theta'_2}$.
\State \textbf{Initialize:} Replay buffer $D$ with capacity $B$.
\For{each episode}
\State Generate random channel gains $h^{(1)}, h^{(2)}$ and get initial state $s$. \Comment{State $s$ from Eq. \ref{eq:observation_space}}
\For{each step of the episode}
\State Select action $a_t \sim \pi_\phi(a|s_t)$. \Comment{Action $a_t = \rho^{(1)}$ from Eq. \ref{eq:action_space}}
\State Execute action $a_t$, calculate throughputs and BER.
\State Observe reward $r_t$ and next state $s_{t+1}$. \Comment{Reward $r_t$ from Eq. \ref{eq:total_reward}}
\State Store transition $(s_t, a_t, r_t, s_{t+1})$ in $D$.
\State Sample a random mini-batch of $M$ transitions $(s_i, a_i, r_i, s'_i)$ from $D$.
\State \Comment{Compute the target for the Q-functions}
\State With next actions $a'_i \sim \pi_\phi(a'|s'_i)$, the target is:
\Statex \hspace{\algorithmicindent} $y_i = r_i + \gamma \left( \min_{j=1,2} Q'_{\theta'_j}(s'_i, a'_i) - \alpha \log \pi_\phi(a'_i|s'_i) \right)$.
\State \Comment{Update the critic networks}
\State Update $\theta_j$ by minimizing the loss: $\mathcal{L}(\theta_j) = \frac{1}{M} \sum_{i} (Q_{\theta_j}(s_i, a_i) - y_i)^2$ for $j=1,2$.
\State \Comment{Update the actor network}
\State With new actions $\tilde{a}_i \sim \pi_\phi(a|s_i)$, update $\phi$ by maximizing the objective:
\Statex \hspace{\algorithmicindent} $J(\phi) = \frac{1}{M} \sum_{i} \left( \min_{j=1,2} Q_{\theta_j}(s_i, \tilde{a}_i) - \alpha \log \pi_\phi(\tilde{a}_i|s_i) \right)$.
\State \Comment{Update the target networks}
\State Soft update target critic weights: $\theta'_j \leftarrow \tau \theta_j + (1 - \tau) \theta'_j$ for $j=1,2$.
\EndFor
\EndFor
\end{algorithmic}
\end{algorithm}

For a 10 dB SNR, a BER of around $10^{-2}$ was achieved for the SAC agent, while the BER graphs obtained for GRPA and NGDPA are shown in Fig.~\ref{fig:6a}. While not error-free at this low SNR, the agent's power allocation is significantly more effective at minimizing errors compared to GRPA and NGDPA. The agent's strategy ensures that the interference levels are managed well enough for the SIC process at the receiver to be highly successful relative to the alternatives, leading to superior communication reliability under the tested conditions. In contrast, the higher BER of GRPA and NGDPA indicates that their static allocation rules can lead to residual interference levels that are too high for reliable decoding.

As shown in Fig.~\ref{fig:5a}, our SAC-based agent maintains a perfect fairness index across the entire range of channel gain ratios. In stark contrast, traditional schemes such as NGDPA suffer significant fairness degradation when users' channel conditions become disparate, and GRPA performance deteriorates as the channel gains become more similar, while NGDPA's fairness improves but at the cost of total throughput. These findings underscore the primary advantage of our adaptive SAC-based approach in ensuring equitable service across diverse channel environments. This consistent fairness is a direct outcome of the fairness centric sophisticated reward function design, which implements a quadratic penalty for fairness deviations and heavily rewards achieving the target fairness threshold. The agent effectively learns to balance maximizing the overall throughput, shown in Fig.~\ref{fig:4a}, with the critical requirement of not starving any single user, which is a common pitfall for purely throughput-driven algorithms. It achieves the fairness level of an orthogonal method like TDMA while providing the superior spectral efficiency of NOMA.

To provide a concrete numerical illustration of these differences, Table \ref{tab:comparison_example} compares the performance of the SAC agent, GRPA, and NGDPA for a specific scenario where the channel gain ratio $r = 0.5$. For this example, we selected a mid-range channel gain for the stronger user, $h^{(1)} = 3.132 \times 10^{-4}$, which results in $h^{(2)} = 1.566 \times 10^{-4}$. The table clearly shows the resulting power allocations, user throughputs, and BER performance, highlighting the practical implications of each strategy.

\begin{table}[h!]
\centering
\caption{Performance comparison for a specific scenario ($r=0.5$, $h^{(1)}=3.132 \times 10^{-4}$)}\label{tab:comparison_example}
\begin{tabular}{@{}lccccc@{}}
\toprule
\textbf{Method} & \textbf{\begin{tabular}[c]{@{}c@{}}Power Alloc. \\ ($\rho^{(1)}$)\end{tabular}} & \textbf{\begin{tabular}[c]{@{}c@{}}Tput $U_1$ \\ (Mbps)\end{tabular}} & \textbf{\begin{tabular}[c]{@{}c@{}}Tput $U_2$ \\ (Mbps)\end{tabular}} & \textbf{BER $U_1$} & \textbf{BER $U_2$} \\
\midrule
SAC Agent & 0.0333 & 102.03 & 102.21 & 1.52e-02 & 9.16e-03 \\
GRPA & 0.2000 & 100.36 & 82.93 & 3.14e-02 & 2.44e-02 \\
NGDPA & 0.3333 & 93.19 & 52.96 & 1.01e-01 & 9.23e-02 \\
\bottomrule
\end{tabular}
\end{table}

\subsection{Computational Complexity Analysis}

A crucial aspect for the practical deployment of any resource allocation scheme is its computational complexity. While the proposed SAC-based approach involves a training phase, this is a one-time, offline cost. The relevant metric for real-time performance is the inference complexity, which is the time taken to make a power allocation decision during operation.

Table \ref{tab:complexity} provides a qualitative comparison of the computational complexity of the SAC agent during inference against the conventional GRPA and NGDPA methods.

\begin{table}[h!]
\centering
\caption{Computational complexity comparison of power allocation schemes}\label{tab:complexity}
\begin{tabular}{@{}lll@{}}
\toprule
\textbf{Method} & \textbf{Operation Type} & \textbf{Real-time Complexity} \\
\midrule
GRPA/NGDPA & \begin{tabular}[c]{@{}l@{}}Direct calculation from\\ channel gain values\end{tabular} & Low ($O(1)$) \\
SAC Agent & \begin{tabular}[c]{@{}l@{}}Forward pass through a\\ pre-trained neural network\end{tabular} & \begin{tabular}[c]{@{}l@{}}Low, depends on network\\ size but constant per decision\end{tabular} \\
\bottomrule
\end{tabular}
\end{table}

As shown in the table, GRPA and NGDPA involve simple analytical calculations with constant low complexity. The SAC agent's decision-making relies on a forward pass through its actor network. Although this is computationally more intensive than a simple formula, modern hardware can perform this operation in sub-millisecond timescales for the network sizes typically used in such applications. Therefore, the marginal increase in computational overhead is a justifiable trade-off for the significant improvements in system performance, fairness, and robustness demonstrated in our simulations. Essentially, we trade a negligible, constant-time inference cost for a dynamic, intelligent allocation policy that adapts to any channel condition, a capability that fixed-rule methods inherently lack.

\subsection{Illumination Analysis}
In order to assess the illumination performance of both the proposed and reference systems, we model the LED's output PSD in two scenarios: without communication (baseline PSD) and under CSK modulation (altered PSD). The modulation process perturbs the average spectral distribution of the emitted light, thereby impacting the illuminance. The power contributions of the transmitted symbols for users $U_1$ and $U_2$ in color channel $i$ are given by Eq.~\ref{eqn:88}. Using these contributions, the modified PSD, denoted $PSD_a(\lambda)$, is computed via Eq.~\ref{eqn:22}, and this altered spectrum is subsequently employed to evaluate color metrics CRI, CCT, and luminous flux.

\begin{equation}
TSP_i=\mathbb{E}\left(\left[S_i^{(1)} {\rho^{(1)}}+S_i^{(2)} {(1-\rho^{(1)})}\right]^2\right)
\label{eqn:88}
\end{equation}
\begin{equation}
PSD_a(\lambda)=\sum_{i=1}^3 PSD_i(\lambda) TSP_i
\label{eqn:22}
\end{equation}

In this case, after obtaining $PSD_a$, CRI, CCT and luminous flux parameters for the cases with and without communication are calculated using the open source program \cite{matlabexchange_pspectro} and given in Table \ref{tab:3}. The NOMA method did not change the CRI and CCT as expected. Only a decrease in luminous flux was observed due to modulation in communication. The stability of the CRI and CCT is a critical result for practical applications, as it ensures that the primary function of lighting is not compromised by the secondary data transmission task. The reduction in luminous flux is an anticipated and fundamental trade-off in VLC, as a portion of the optical power is utilized to carry data rather than solely for illumination.

\begin{table}[h!]
\centering
\caption{Illuminations parameters comparison table}\label{tab:3}
\begin{tabular}{@{}cccc@{}}
\toprule
& CRI & CCT & Luminous flux (lm) \\
\midrule
W/O Comm. & 29.87 & 10105 & 5713 \\
W Comm. CSK Only & 29.84 & 10105 & 1714 \\
W Comm. NOMA CSK & 29.84 & 10105 & 1714 \\
\bottomrule
\end{tabular}
\end{table}

\subsection{SNR-BER Analysis}
\begin{figure}[H]
\centering
\begin{subfigure}[b]{0.46\textwidth}
\includegraphics[width=\textwidth]{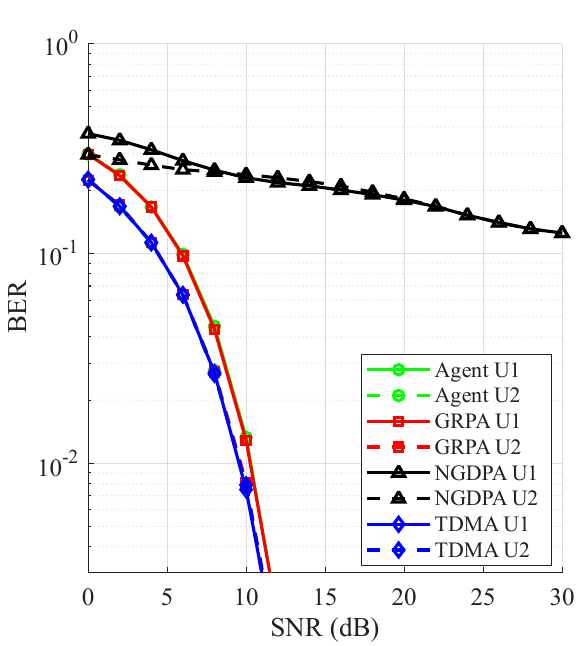}
\caption{SNR vs. BER at $h_2/h_1 = 0.10$}
\label{fig:7a}
\end{subfigure}
\hfill
\begin{subfigure}[b]{0.46\textwidth}
\includegraphics[width=\textwidth]{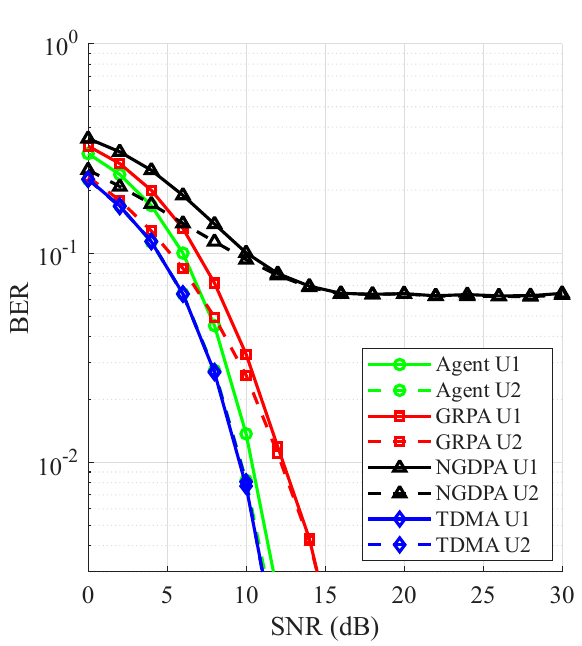}
\caption{SNR vs. BER at $h_2/h_1 = 0.50$}
\label{fig:8a}
\end{subfigure}
\vskip\baselineskip
\begin{subfigure}[b]{0.46\textwidth}
\includegraphics[width=\textwidth]{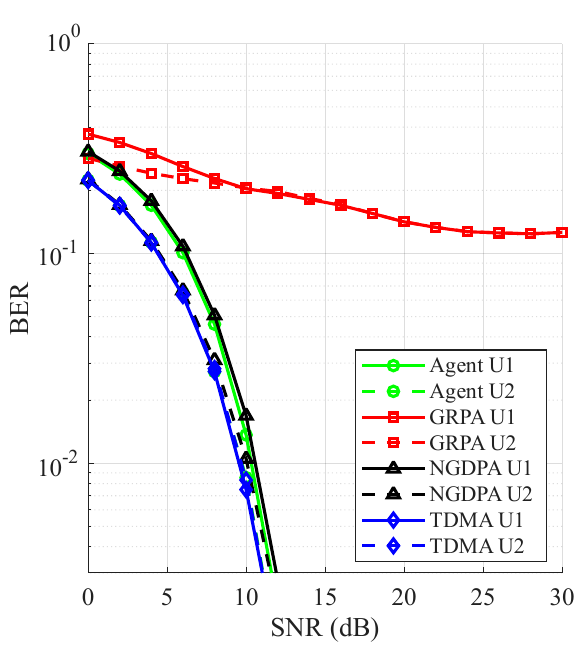}
\caption{SNR vs. BER at $h_2/h_1 = 0.90$}
\label{fig:9a}
\end{subfigure}
\caption{BER performance comparison versus SNR for different channel ratios.}
\label{fig:7-10}
\end{figure}

To further analyze the performance under different conditions, we evaluated the BER as a function of SNR for three distinct channel ratios ($h^{(2)}/h^{(1)} \in \{0.1, 0.5, 0.9\}$).  For this example, we selected a mid-range channel gain for the stronger user, $h^{(1)} = 3.132 \times 10^{-4}$ as before. The results are presented in Fig.~\ref{fig:7-10}. Across all evaluated scenarios, the proposed SAC agent consistently outperforms both GRPA and NGDPA, achieving lower BER for any given SNR. Notably, under high channel gain similarity (e.g., $r = 0.9$), where user conditions are nearly uniform, GRPA exhibits the poorest performance, with the BER plateauing around $10^{-1}$ at 30~dB. Conversely, at lower channel gain ratios, NGDPA demonstrates the worst BER, rendering it impractical for reliable communication. This comprehensive analysis underscores the robustness of the SAC agent’s learned policy, which maintains superior performance across diverse channel conditions and a broad range of SNR levels.

\begin{figure}[H]
\centering
\begin{subfigure}[b]{0.49\textwidth}
\includegraphics[width=\textwidth]{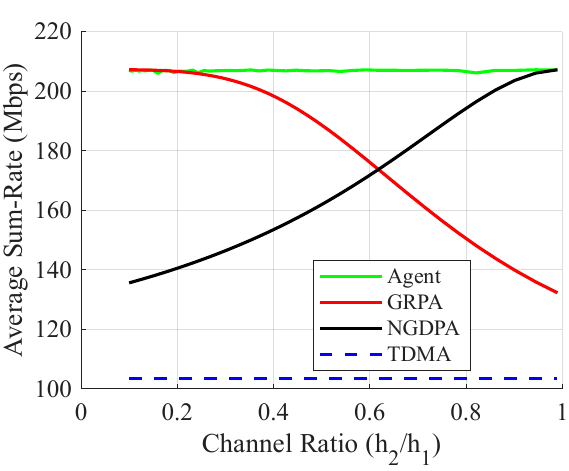}
\caption{Sum-rate performance}
\label{fig:11a}
\end{subfigure}
\hfill
\begin{subfigure}[b]{0.49\textwidth}
\includegraphics[width=\textwidth]{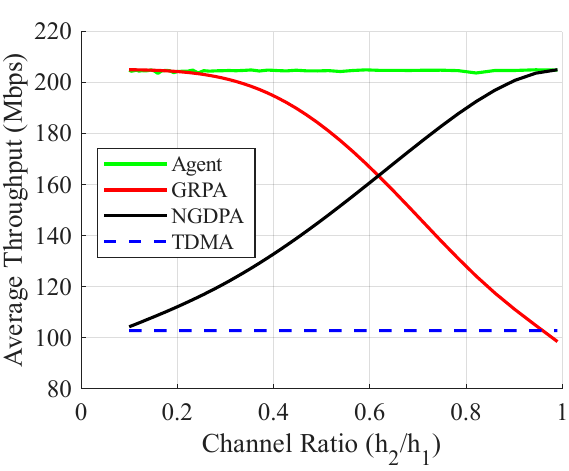}
\caption{Throughput performance}
\label{fig:12a}
\end{subfigure}
\vskip\baselineskip
\begin{subfigure}[b]{0.49\textwidth}
\includegraphics[width=\textwidth]{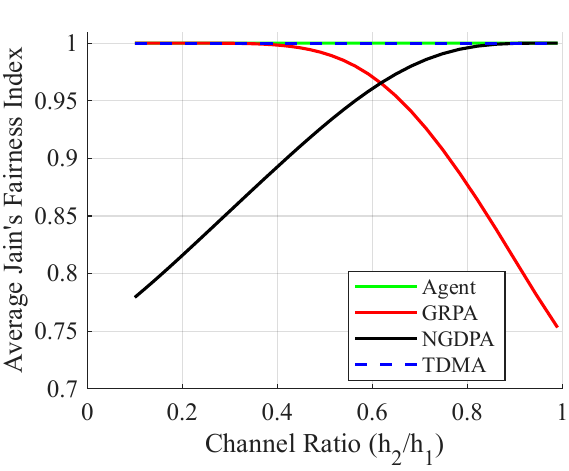}
\caption{Fairness performance}
\label{fig:13a}
\end{subfigure}
\hfill
\begin{subfigure}[b]{0.49\textwidth}
\includegraphics[width=\textwidth]{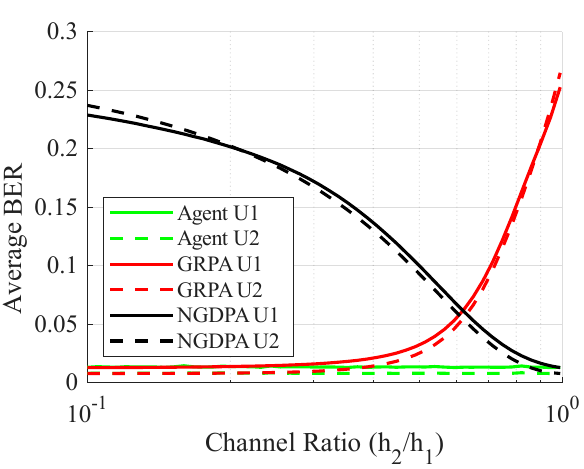}
\caption{BER performance}
\label{fig:14a}
\end{subfigure}
\caption{Performance comparison of the proposed SAC agent against GRPA, NGDPA, and TDMA outside the trained environment.}
\label{fig:11-14}
\end{figure}

\subsection{Adaptability and Generalization in Unseen Environment}

The agent was trained for a specific channel range. To investigate its performance in an environment other than where it was trained, we tested it in a new environment with a lower bound value 5 times smaller and an upper bound value 5 times larger than that of the environment it was trained in. We compared the agent's performance in similar way we did in trained environment. As can be seen from Fig.~\ref{fig:11a}, Fig.~\ref{fig:12a}, and Fig.~\ref{fig:13a}, the agent performs better in terms of sum-rate, throughput, and fairness compared to other methods in an environment outside of where it was trained. The BER graph for NGDPA and GRPA is given in Fig.~\ref{fig:14a} while a significantly lower BER is achieved with the SAC agent. This set of results is arguably the most significant contribution of this work. The consistent and superior performance of the SAC agent in this "unseen" environment, as depicted in Figure \ref{fig:11-14}, demonstrates its remarkable ability to generalize. It implies that the agent has not merely memorized the optimal responses for the training data but has successfully learned an underlying, robust control policy that maps channel-state information to an effective power allocation strategy. This generalization capability is crucial for practical deployment, as it obviates the need for frequent retraining in dynamic real-world scenarios, making the proposed solution both powerful and scalable.

\section{Conclusion}

In this paper, we introduce and analyze a novel VLC system architecture integrating NOMA-CSK. Addressing the critical challenge of resource management within this framework, we developed a sophisticated power allocation solution using a SAC reinforcement learning agent. Our primary objective was to maximize user fairness, a crucial requirement for Quality of Service in modern wireless networks.

Our work delivers two key outcomes. First, we established the viability and modeled the performance of the NOMA-CSK concept. Second, we demonstrate that our SAC-based agent is a highly effective method for managing power allocation within it. The agent consistently and substantially surpassed traditional GRPA and NGDPA methods in fairness, sum-rate, and error performance. The most impactful finding is the agent's remarkable ability to generalize: It performed optimally in new, unseen environments with a significantly wider range of channel conditions, without any retraining, proving it has learned a truly adaptive policy. This robustness underscores the practicality of the RL-based approach for real-world wireless systems where conditions are inherently unpredictable. Furthermore, our analysis of illumination parameters confirmed that the proposed system maintains lighting metrics such as CRI and CCT, and also provides the same level of illumination as CSK only system ensuring stability for practical indoor environments.

While the SAC agent introduces a minor computational overhead during real-time operation compared to simpler heuristics, we argue this is a negligible trade-off for the substantial gains in fairness, spectral efficiency, and robustness. This research provides a framework for NOMA-CSK systems and offers a powerful, intelligent solution to one of its core operational challenges. Future work will focus on scaling this framework to include more users and testing its resilience under NLoS conditions to further pave the way for next-generation, high-efficiency indoor wireless networks.

\section*{Author Contributions}

S.V. conceptualized the study, performed the simulations, analyzed the data, and wrote the original manuscript. G.H. supervised the project, provided critical guidance and feedback, and edited the manuscript. Both authors reviewed and approved the final version.

\section*{Funding}

This research received no external funding.

\section*{Data Availability}

Data sharing is not applicable to this article.

\section*{Declarations}
\textbf{Conflict of interest:} The authors declare that they have no conflict of interest. \\
\textbf{Consent to participate:} The authors consent to participate. \\
\textbf{Ethical approval:} Not applicable. \\
\textbf{Consent for publication:} The authors give their consent for publication.

\end{document}